\documentstyle[prd,preprint,aps,epsfig,floats]{revtex}
\draft
\newcommand{\beq}{\begin{equation}}
\newcommand{\eeq}{\end{equation}}
\newcommand{\bea}{\begin{eqnarray}}
\newcommand{\eea}{\end{eqnarray}}
\begin{document}

\title{Diffractive Exclusive Photon Production in DIS at HERA} 

\author{L.L.Frankfurt$^a$, A.Freund$^b$, M. Strikman$^b$}

\address{
$^a$Physics Department, Tel Aviv University, Tel Aviv, Israel\\ 
$^b$Department of Physics, Penn State University\\
University Park, PA  16802, U.S.A.}

\maketitle

\begin{abstract}
We demonstrate that perturbative QCD allows one to calculate the absolute
cross section of diffractive exclusive production of photons
at large $Q^2$ at HERA, while the aligned jet model allows one to estimate 
the cross section for intermediate $Q^2 \sim 2 GeV^2$.  
 Furthermore, we find that the imaginary part of the amplitude for the 
production of real photons is larger than the imaginary part of the
corresponding  DIS amplitude by about a factor of $2$, leading to the 
prediction of a significant 
counting rate for the current generation of experiments at HERA. We also find
a large azimuthal angle asymmetry 
in $ep$ scattering for
HERA kinematics which allows one to directly measure the real part of the 
DVCS amplitude and hence the nondiagonal parton distributions.
\end{abstract}

\section{Introduction}
\label{sec:intro}

Recent data from HERA has spurred great interest in exclusive or diffractive 
direct production of photons  in $e - p$ scattering (DVCS-
deeply virtual Compton scattering) as another source to obtain more 
information about the gluon distribution inside the proton for nonforward 
scattering. In recent years studies of diffractive vector meson production and
deeply virtual Compton scattering has greatly increased
our theoretical understanding about the gluon distribution in 
nonforward kinematics and how it compares to the gluon distribution in the 
forward direction. For a less than complete list of recent
references see Ref.\ \cite{1,2,3,4,ours,6,7,8,9,MR97,av}.

Exclusive diffractive virtual Compton processes at large $Q^2$,
first investigated in \cite{10}, offer a new and 
comparatively ``clean''\cite{f1}
way of obtaining information about the gluons inside the proton in a 
nonforward kinematic situation. We are interested in the production of a 
real photon compared to the inclusive DIS cross section. The
 exclusive process is nonforward 
in its nature, since the photon initiating the process is virtual 
($q^2<0$) and the final state photon is real, forcing a small but finite
momentum transfer to the target proton i.e forcing a nonforward kinematic 
situation as we would like. 
We will show that pQCD can be applied to this type of  exclusive
process although we will not give a formal proof on the level 
comparable to the DIS case \cite{Collins}. Such a proof can be found in 
Ref.\ \cite{ca}.

The paper is organized in the following way.
In Sec.\ \ref{AJM} we estimate the amplitude in the normalization point
$Q_0^2 \sim 2 GeV^2$ using
the aligned jet model approximation and conclude 
that for such $Q^2$ the nondiagonal amplitude is larger than the diagonal one 
by a factor of $\sim 2$.
 In Sec.\ \ref{sec:hp} we 
calculate the imaginary part of the amplitude for $\gamma^* + p \rightarrow 
\gamma + p$ in the leading order
of the running coupling constant $\alpha_s$ and compare it to the 
imaginary part of the amplitude in DIS in the same order. 
In Sec.\ \ref{Bslope} we argue that at sufficiently small $x$
the  $t$-dependence of the cross section should reflect the interplay
of hard and soft physics typical of diffractive phenomena in
DIS. Namely, that 
at fixed $x$ and increasing $Q^2$, hard physics should tend to occupy
the dominant part of the space of rapidities. In contrast to this, at 
fixed $Q^2$
and decreasing $x$, hard physics should occupy a finite 
range of rapidities which increases with $Q^2$ - 
  $~\sim \ln {Q^2\over \beta M_{\rho}^2} $
with $\beta\sim  0.1 - 0.2 $ at the HERA energy range
due to the QCD evolution, and that soft QCD physics occupies the rest of the 
phase space. 
 In Sec.\ \ref{sec:slope} we give the total cross section of exclusive photon 
production and give numerical estimates of the DVCS production rate 
at HERA and find that such measurements are feasible for the current 
generation of experiments. We also show the feasibilty of directly measuring 
the real part of the DVCS amplitude and hence, at least, the shape of the 
nondiagonal parton distributions through a large azimuthal angle asymmetry in 
$ep$ scattering for HERA kinematics.
Sec.\ \ref{sec:concl} finally contains concluding remarks.

\section{The amplitude for diffractive virtual Compton scattering at 
intermediate $Q^2$ }
\label{AJM}
Similar to the case of deep inelastic scattering, in real photon
production it is possible to calculate within perturbative QCD
the $Q^2$ evolution of the amplitude but not its value at the normalization
point at $Q_0^2 \sim $ {\it few GeV$^2$}
where it is given by nonperturbative effects. 
Hence we start by discussing expectations for this region. It was demonstrated 
in \cite{FS88} that the aligned jet model \cite{bj} coupled with the idea of 
color screening provides a reasonable semiquantitative description of
$F_{2N}(x \le 10^{-2}, Q_0^2)$. 
In this model the virtual photon interacts at intermediate $Q^2$ and small $x$
via transitions to a $q \bar q$ pair with small transverse momenta - 
$k_{0,t}$ ($\left<k_{0,t}^2\right> \sim 0.15 GeV^2$)
and average masses $\sim Q^2$ which thus carry asymmetric fractions of the
virtual photon's longitudinal momentum. Due to large transverse color
separation, 
$b \sim 2\sqrt {2/3} r_{\pi}$,
  the aligned jet model components of the photon wave 
function interact strongly with the target with the cross section
$\sigma_{tot}(``AJM''-N) \approx \sigma_{tot}(\pi N)$.
Neglecting contributions of the components of the
$\gamma^*$ wave function with smaller color
separation, one can write $\sigma_{tot}(\gamma^*N)$
 using the Gribov dispersion representation \cite{Gribov} as \cite{FS88}:
\beq
\sigma_{tot}(\gamma^*N)= {\alpha \over 3 \pi}
 \int_{M_0^2}^{\infty}{\sigma_{tot}(``AJM''-N) R^{e^+e^-}(M^2)
M^2 {3 \left<k_{0~t}^2\right>\over M^2} \over (Q^2+M^2)^2}d M^2,
\label{AJMeq}
\eeq
where the factor $M^2$ in the  nominator is due to the 
overall phase volume, 
$R^{e^+e^-}(M^2)={\sigma(e^+e^- \to hadrons) \over 
\sigma(e^+e^- \to \mu^+\mu^-)}$. The factor 
${3 \left<k_{0~t}^2\right>\over M^2}$ is the fraction of the whole phase 
volume occupied by the aligned jet model ,
and the factor $1/(Q^2+M^2)^2$ is due to 
the propagators of the photon in the hadronic intermediate state
with mass$^2$ equal $M^2$. Based on the logic of 
local quark-hadron  duality (see e.g.\cite{FRS} and references
therein) we take the
lower limit of integration $M_0^2 \sim 0.5 GeV^2 \le m_{\rho}^2$.
  In the case of real photon production the imaginary part of the 
amplitude for $t=0$ is 
\beq
{1\over s} Im A(\gamma^*+N \to \gamma +N)_{t=0}= {\alpha \over 3 \pi}
\int_{M_0^2}^{\infty}
{\sigma_{tot}(``AJM''-N) R^{e^+e^-}(M^2)
M^2 {3 \left<k_{0~t}^2\right>\over M^2} \over (Q^2+M^2)M^2}d M^2,
\label{ImAJMeq}
\eeq 
with $s=2 q_0m_N$ being the flux factor.
The only difference from Eq.\ \ref{AJMeq} for
$\sigma_{tot}(\gamma^*+N)$ is 
the change of one of the propagators from $1/(Q^2+M^2)$ to $1/M^2$ -
here $q_0$ is the energy of the virtual photon in the rest frame of the 
target.
 
Approximating $R^{e^+e^-}(M^2)$ 
as a constant for the $Q^2$ range in question (we understand this in 
the sense of a local duality of the hadron spectrum and 
the $q \bar q$ loop) we find 
\beq
R \equiv {Im A(\gamma^*+N \to \gamma^* +N)_{t=0}
\over Im A(\gamma^*+N \to \gamma +N)_{t=0}}= {Q^2 \over Q^2+M_0^2}
\ln^{-1}(1+Q^2/m_0^2).
\label{ratiogam}
\eeq
In the following analysis we will take
$Q^2_0$ for the
perturbative QCD evolution as 2.6 GeV$^2$ to avoid ambiguities. 
It is easy convince oneself that for $M_0^2 \sim 0.4 \div 0.6~ GeV^2$  and 
$Q^2\approx 2-3~\mbox{GeV}^2$ Eq.\ref{ratiogam} leads to $R \approx 0.5$.
A similar value of $R$ has been obtained within the generalized
vector dominance model \cite{FGS}
 As we will see below QCD evolution leads to a strong increase of
$Q^2 Im A(\gamma^*+N \to \gamma +N)_{t=0} $ with increase of $Q^2$
for fixed $x$. However it does not change appreciably the value of $R$.

\section{The amplitude for exclusive real photon production at large $Q^2$.}
\label{sec:hp}

\begin{figure}
\centering
\mbox{\epsfig{file=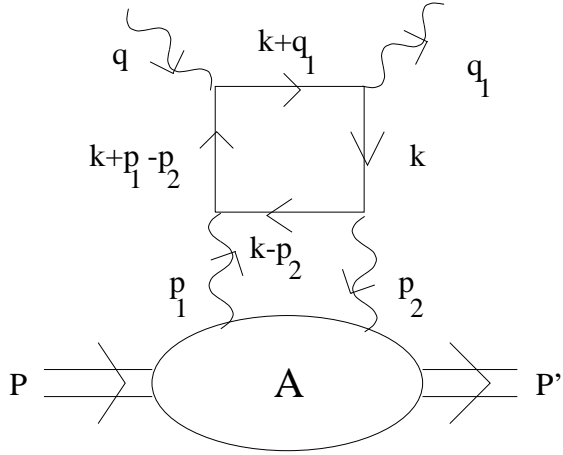,height=6cm}}
\vspace*{5mm}
\caption{Leading contribution to DVCS at small $x$.}
\label{dem}
\end{figure}

The process of exclusive direct production of photons in 
first nontrivial order of $\alpha_s \ln {Q^2\over Q_0^2}$
at small $x_{Bj}$ can be calculated (see Fig.\ \ref{dem}) as the sum of a 
hard contribution 
calculated within the framework of QCD evolution equations \cite {Abramowicz}
and a soft contribution which we evaluated above within the aligned jet model. 
The hard contribution can be described through a two gluon exchange of a box 
diagram with the target proton.
In order to calculate the imaginary part of the amplitude, we need to 
calculate the hard amplitude from the box
as well as the gluon-nucleon scattering plus the 
soft aligned jet model contribution. 
Let us first give a general 
expression for the imaginary part of the amplitude and then proceed to deal 
with the gluon-nucleon scattering, followed by the calculation of the box 
diagrams. 

First, let us discuss the hard contribution which actually dominates
in the considered process. To account for the gluon-nucleon scattering, 
we work with Sudakov variables for the gluons with momenta $p_1$ and $p_2$
 attaching the box to the target and the following kinematics  for the 
gluon-nucleon scattering:
\beq
p_1 = \alpha q' + x_1 p' + p_{t}, \; \; d^{4}p_1=\frac{s}{2}d\alpha dx_1 d^{2}
p_{t},
\label{suva}
\eeq
where $q'$ and $p'$ are light-like momenta related to $p,q$ the momenta of the
target proton and the probing virtual photon respectively, by:
\bea
& &q = q' - xp', \; \; p = p' + \frac{p^2}{2p'q'}q', \nonumber\\
& &s = 2pq = 2p'q'- xp^2,
\eea
with $x$ being the Bjorken $x$ and $x_1$ the proton momentum fraction 
carried by the outgoing gluon. 
Equivalent equations to Eq.\ \ref{suva} apply 
for $p_2$ with the only difference being that $x_1$ is replaced by $x_2$,
the momentum fraction of the incoming gluon,
signaling that there is only a difference in the longitudinal momenta but not 
in the transverse momenta. This fact will shortly become important. 
Furthermore there is a simple relationship between $x_1$ and $x_2$ : 
$\Delta = x_1 - x_2 = \mbox{const.}$ following from the kinematics of the 
considered reaction\cite{f2} where $\Delta$ is the asymmetry parameter or 
skewedness of the process under consideration.
Therefore one is left with just 
the integration over $p_1$ since $p_2$ cannot vary independently of $p_1$.  
Being exclusively interested in the small $x$ region, one can safely make the 
following approximations: $s = 2pq \simeq 2p'q'$ and $p' \simeq p$. Since we 
are working in the leading $\alpha_s \ln Q^2$ approximation, neglecting 
corrections of order $\alpha _s$, the main contribution comes from the region 
$p_{t}^2<<Q^2$, hence the contribution to the imaginary part of the amplitude 
simplifies considerably. First, since 
$|p^{2}_{1}|=|\alpha x_1 s + p_{t}^2| << Q^2$, one has $\alpha << 1$ and the 
polarization tensor of the 
propagator of the exchanged gluon in the light-cone gauge $q'_{\mu}A^{\mu}=0$ 
becomes, see Ref.\ \cite{Gribov} :
\beq
d_{\mu\lambda}\simeq \frac{p'_{\mu}q'_{\lambda}}{p'q'}.
\label{prop}
\eeq
In other words it is enough to take the longitudinal polarizations of the 
exchanged gluons into account.

Using Eq.\ \ref{prop} one obtains the following expression for the total 
contribution of the box diagram and its permutations:
\beq
Im\, A = \int \frac{d^{4}p_1}{(2\pi)^{4}i}\frac{1}{p^2_1p^2_2}
2ImA^{ab(P)}_{\mu\nu}ImA^{ab(T)}_{\lambda\sigma}d_{\mu\lambda}(p_1)
d_{\nu\sigma}(p_2),
\label{amp}  
\eeq
where $ImA^{ab(P)}_{\mu\nu}=ImA^{ab}_{\mu \nu}(\gamma^* g 
\rightarrow q\bar q)$ is 
the sum of the box diagrams, $ImA^{ab(T)}_{\lambda \sigma}$ is the amplitude 
for the gluon-nucleon scattering, a,b  are the color indices and the 
overall tensor structure has been neglected for now. The usage of
the imaginary part of the scattering amplitude and in particular limiting
ourselves to the s-channel contribution as the dominant part in both
the forward and the nonforward case (Eq.\ \ref{amp}) is correct 
(see Ref.\ \cite{ours} for more details) as long as we restrict ourselves to 
the DGLAP region of small $x$ and thus small $t$, where 
 $t=(p_1-p_2)^2$ is the square of the momentum transfered to the target.
The real part of the amplitude will be evaluated below by applying a
dispersion relation over the center of mass energy $s$.
Using Eq.\ \ref{prop} and the Ward identity which is the same as in the 
Abelian case since the box contains no gluons i.e is color neutral: 
\beq
A^{ab(P)}_{\mu \nu} p_{1\mu } = 0,
A^{ab(P)}_{\mu \nu} p_{2\nu} = 0,
\eeq
yielding
\beq
\frac{ImA^{ab(P)}_{\mu \nu} p^{'}_{\mu} p^{'}_{\nu}}{4(pq)^{2}} = 
\frac{ImA^{ab(P)}_{\mu \nu} p_{t\mu} p_{t\nu}}{x_{1}x_{2}s^{2}},
\eeq
one can rewrite Eq.\ \ref{amp} as:
\beq
{Im A\over s}= \int^{1}_{x} \frac{dx_1}{x_1} 
E(x/x_1,\Delta/x_1,Q^2,p_t^2,Q_0^2)\int \frac{s d\alpha d^{2}p_{t}}
{(2\pi )^{4} p_{1}^{2} p_{2}^{2}} p_{t}^{2} \Sigma _{a} 
\frac{4ImA^{a(T)}_{\lambda \sigma} q_{\lambda} q_{\sigma}}{s^2},
\label{iamp}
\eeq
where we have used $<p_{t\, \mu}p_{t\, \nu}>=-\frac{1}{2}g_{\mu\nu}^{t}
p^{2}_{t}$ (the average over the transverse gluon polarization) and defined 
the imaginary part of the hard scattering to be given by:
\beq
E(x/x_1,\Delta/x_1,Q^2,p_t^2,Q_0^2)= - \frac{1}{2}g_{\mu\nu}^{t} 
\frac{ImA_{\mu\nu}^{ab(P)}}{x_2 s}\delta_{ab},
\eeq
where the sum over repeated indices is implied. Up to this point we have just
rewritten the equation for the imaginary part of the total 
amplitude but
have not identified the different perturbative and non-perturbative pieces.
In the case of a virtual photon with longitudinal polarization, this would be 
an easy task since the $q\bar q$ pair would only have a small space-time 
separation and we could follow the argument in Ref.\ \cite{1,4,FRS} stating
that the box is entirely dominated by the hard scale $Q$ and thus can 
unambiguously be calculated in pQCD. However, in our case we are dealing with 
a virtual photon which is transversely polarized and thus one can have large
transverse space separations between $q$ and $\bar q$. The resolution
to this problem can be found in the following way: one accepts that 
one has a contribution from a soft, aligned-jet-model-type, configuration
and  that there is no unambiguous separation of the amplitude in a 
perturbative and non-perturbative part up to a certain scale $Q_0^2$.
However, in the integration over transverse gluon momenta, one
will reach a scale at which a clear separation into perturbative and 
non-perturbative part can be made and hence we can unambiguously
calculate albeit not the imaginary part of the amplitude of the 
upper box but its $\ln Q^2$ derivative i.e its kernel convoluted with a parton
distribution. At this point then, one can 
include the non-perturbative contribution of the aligned jet model
into the initial 
distribution of the imaginary part of the total amplitude and solve the 
differential equation in $Q^2$. One obtains the following solution for the 
imaginary part \cite{Abramowicz}:
\beq
ImA(x,Q^2,Q_0^2) = ImA(x,Q_0^2)+
\int^{Q^2}_{Q_0^2}{dQ'^{2}\over Q'^{2}} 
\int^{1}_{x} \frac{dx_1}{x_1}P_{qg}(x/x_1,\Delta/x_1)g(x_1,x_2,Q'^2),
\label{tamp}
\eeq
where $P_{qg}$ is the evolution kernel\cite{f3} and starting from $Q_0^2$
the gluon distribution can be defined from 
Feynman diagrams in the leading $\alpha _s \ln Q^2$ approximation by 
realizing that in Eq.\ \ref{iamp} one can replace 
$p_{1}^{2}$ and $ p_{2}^{2}$ by $p_{t}^{2}$ and one finds:
\beq
\int \frac{s d\alpha d^{2} p_{t}}{(2\pi )^{4} p_{t}^2} 
\Sigma _{a} 
\frac{4ImA^{a(T)}_{\lambda \sigma} q_{\lambda} q_{\sigma}}{s^2} = 
g(x_{1},x_2,Q^2),
\label{pdis}
\eeq
where $g$ is the nondiagonal parton distribution in general.
Comparison of Eq.\ \ref{iamp} with the QCD-improved parton
model expression for the total cross section of charm production given in 
\cite{12} shows that $g$ in the case $\Delta=0$
is the conventional, diagonal gluon distribution.

Note that the parton distribution which
serves as an input in Eq.\ \ref{tamp} has to be evolved over the $Q^2$-range
covered by the $Q'^2$ integral which complicates the calculation. We will
explain below how to deal with this issue in practical situations.

At this point we would like to comment on equivalent definitions 
of nondiagonal parton distributions in the literature which 
differ by kinematic factors (see for example \cite{4,6,7}). 
Eq.\ \ref{pdis} corresponds to the definition used in \cite{7}, however since 
it is given on the level of Feynman diagrams there are no ambiguities
such as renormalization of bilocal operators and hence it provides 
an unambiguous definition of a nondiagonal parton distribution.

For the  non-perturbative input, $Im A(x,Q_0^2)$
we will be able to use the aligned jet model analysis of Sec.\ \ref{AJM}
and the standard relation between $Im A^{\gamma^*p \to \gamma^*p}(x,Q^2,t=0)$
and $F_{2p}(x,Q^2)$:
\beq
Im A^{\gamma^*p \to \gamma^*p}(x,Q^2,t=0)
=\frac{F_{2p}(x,Q^2)}{4\pi^2\alpha x}.
\eeq

Following the discussion above, we now only need to calculate $P_{qg}$ to 
leading logarithmic accuracy, in order to make predictions for the 
imaginary part of the whole amplitude. Therefore, let us now consider the 
box diagram where the two horizontal quark propagators are cut, corresponding 
to the DGLAP region i.e neglecting the u-channel contribution.

\begin{figure}
\centering
\mbox{\epsfig{file=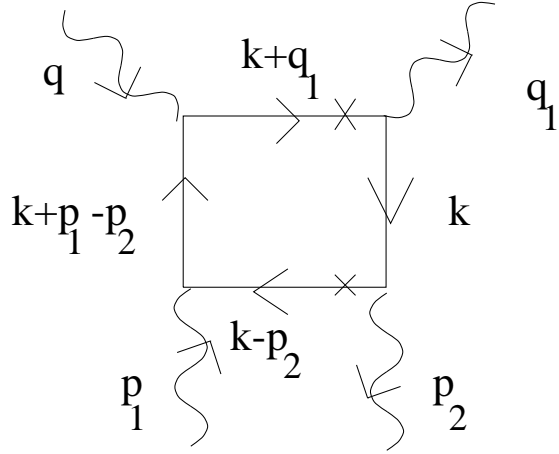,height=6cm}}
\vspace*{5mm}
\caption{Cut box diagram giving the kernel for the imaginary part of the DVCS 
amplitude.}
\label{dem1}
\end{figure}

The kinematics (see Fig.\ \ref{dem1}) for the calculation of the cut box 
diagram, using Sudakov 
variables, is the following. The quark-loop momentum $k$ is given by:
\beq
k = \alpha q' + \beta p' + p_{t}, \; \; d^{4}k=\frac{\hat s}{2}d\alpha d\beta 
d^{2}k_{t},
\eeq
where $q'$ and $p'$ are light-like momenta related to $p,q$ by:
\bea
& &q = q' - xp', \; \; p_1 = p' + \frac{p^2}{2p'q'}q', \nonumber\\
& &\hat s = 2p_1q = 2p'q'- xp_{1}^2.
\eea
The momenta of the exchanged gluons, in light cone coordinates, are given by:
\beq
p_1=(x_1p_+,0,p_t),\; \; p_2=(x_2p_+,0,p_t),
\eeq
where we have assumed the transverse momentum of the proton to be zero.
The probing transverse photon and the produced photon have the following 
momenta, again in light cone coordinates:
\beq
q=(-xp_+,\frac{Q^2}{2xp_+},0_t),\; \;
 q_1=(\simeq 0,\frac{Q^2}{2xp_+},0_t).
\eeq    
$P_{qg}$ is calculated in the light cone gauge yielding the following result 
for the most general case\cite{f4}:
\beq
P_{qg}(x/x_1,\Delta/x_1) = 4 \pi^2 \alpha \frac{\alpha_s}{\pi} 
N_F\frac{
x(x-\Delta) + (x_1-x)^2} {x_1(x_1 - \Delta)^2}. 
\label{cross}
\eeq
The DIS kernel is analogous to Eq.\ \ref{cross} except that $\Delta = 0$
and the kernel for real photon production is obtained for $\Delta=x$.

We now can proceed to calculate the total imaginary part of the amplitude 
from Eq.\ \ref{tamp} where we parameterize the gluon distribution at small 
$x$ as:
\beq
g(x_1,x_2,Q^2)= A_0(Q^2) x_1^{A_1(Q^2)}.
\label{gluondis}
\eeq
We neglect the $x_2$ dependence for the moment\cite{f5}. The above 
parameterization is taken from CTEQ3L as well as the parameterization 
of $\alpha$ in terms of $Q^2$ in leading order:
\bea
A_0(Q^2)&=&exp[-0.7631 - 0.7241\ln\left ( \frac{\ln(Q/\Lambda)}{\ln(Q_0/\Lambda)}
\right ) - 1.17\ln^2\left ( \frac{\ln(Q/\Lambda)}{\ln(Q_0/\Lambda)} \right )
\nonumber\\
& &+ 0.534\ln^3\left ( \frac{\ln(Q/\Lambda)}{\ln(Q_0/\Lambda)}\right )]
\nonumber\\
A_1(Q^2) &=& -0.3573 + 0.3469\ln\left ( \frac{\ln(Q/\Lambda)}
{\ln(Q_0/\Lambda)} \right ) - 0.3396\ln^2\left ( \frac{\ln(Q/
\Lambda)}{\ln(Q_0/\Lambda)} \right )\nonumber \\
& &+ 0.09188\ln^3\left ( \frac{\ln(Q/\Lambda)}{\ln(Q_0/\Lambda)} 
\right ),
\eea
with $\Lambda$, $Q_0$ and $\alpha_s$ given by:
\beq
\Lambda = 0.177\, \mbox{GeV}\;\; Q_0=1.6\,\mbox{GeV}\;\; \alpha_s = 
\frac{4\pi}{9\ln(Q^2/\Lambda^2)},
\eeq
where we have taken $N_C=3$ and $N_F=3$. 

The ratio $R$ of the 
imaginary parts of the amplitudes\cite{f6} is given by:
\beq
R = \frac{ImA(\gamma^* + p \rightarrow \gamma^* + p)}
{ImA(\gamma^* + p \rightarrow \gamma + p)}.
\label{ratio}
\eeq
We give $R$ in the $x$ 
range from $10^{-4}$ to $10^{-2}$ and for a $Q^2$ of $3.5,12$ and $45\; 
\mbox{GeV}^2$ since this kinematic range 
is relevant at HERA. One  might ask what about the contributions due to quarks.
The answer is that the corrections are small\cite{f7}
but for completeness we include 
them here. Eq.\ \ref{tamp} is then augumented with a similar 
expression for the quark contribution where the kernel is now that of 
quark-quark splitting and the 
nondiagonal parton distribution is that of the quark:
\bea
ImA(x,Q^2,Q_0^2)&=& ImA(x,Q_0^2) + \int_{Q_0^2}^{Q^2}
\frac{dQ'^2}{Q'^2}
\int^{1}_{x}\frac{dx_1}{x_1}[P_{qg}(x/x_1,\Delta /x_1)
g(x_1,x_2,Q'^2)\nonumber\\
& & + P_{qq}(x/x_1,\Delta /x_1)q(x_1,x_2,Q'^2)],
\label{conv}
\eea 
where the general expression for the kernel, after a similar calculation as 
before, is found to be:
\beq
P_{qq}(x/x_1,\Delta/x_1)=4\pi^2\alpha\frac{\alpha_s}{\pi}C_F
\left[ \frac{x/x_1 - x^3/x_1^3 - \Delta/x_1\left(x/x_1 + x^2/x_1^2)
\right )}{x_1(1-\Delta/x_1)(1-x/x_1)_+}\right ],
\eeq
and the + - prescription is the one used in Ref.\ \cite{ours}. The quark 
distribution itself is also taken from CTEQ3L\cite{f8} and given by:
\beq
q(x_1,x_2,Q^2)=A_0(Q^2)x_1^{A_1},
\label{qdis}
\eeq
with
\bea
A_0(Q^2)&=&exp[0.1907 + 0.04205\ln\left ( \frac{\ln(Q/\Lambda)}{\ln(Q_0/\Lambda)} 
\right ) + 0.2752\ln^2\left ( \frac{\ln(Q/\Lambda)}{\ln(Q_0/\Lambda)}\right )
\nonumber\\
& &-0.3171\ln^3\left ( \frac{\ln(Q/\Lambda)}{\ln(Q_0/\Lambda)}\right )]
\nonumber\\
A_1&=&0.465.
\eea
We chose $A_1$ to be constant since it varies only between $0.4611$ and $0.468$
in the $Q^2$ range of interest, i.e.\, the error we make is almost negligible
since the quark distribution themselves are small in the $x$-range 
considered. 
Furthermore, according to our discussion in Sec.\ \ref{AJM}, we 
chose the initial distribution for the imaginary part of the DVCS amplitude 
to be twice that of the initial distribution for the imaginary part of the DIS
amplitude.  
In the evolved QCD part, the nonforward kinamtics are taken into account in 
the  kernels of the QCD evolution equation, also the different $Q^2$ 
evolution of nondiagonal as compared to diagonal distribution has been taken 
into account as explained below.

As the calculation with MATHEMATICA showed, the amplitude of the production
of real photons is larger than the DIS amplitude over the whole 
range of small $x$ and $R$ turns out to be between $0.551$, $0.573$ 
and $0.57$ for $x=10^{-4}$, $0.541$, $0.562$ and $0.557$ for $x=10^{-3}$ and 
$0.518$, $0.519$ and $0.505$ for $x=10^{-2}$ in the given $Q^2$ range. 
It has to be pointed out that for a given $Q^2$, the ratio is basically 
constant. Of course, the ratio $R$ will approach $1/2$ as $Q^2$ is 
decreased to the nonperturbative scale since this is 
our aligned jet model estimate
The reason for the deviation from $R=1/2$ is due to the difference in the
evolution kernels.

It is worth noting that in the kinematics we discuss, 
the ratio is still rather sensitive to the nonperturbative boundary
condition. For example, assuming the same boundary
conditions for DVCS and DIS, would result in  a reduction  of $R$ of about 
$20(10)\%$ at $Q^2 \sim 12 (40)~ \mbox{GeV}^2$ and $x \sim 10^{-3}$

In Eq. \ref{tamp} the median point of the integral corresponds to
$x_1/2 \sim x_2 \approx x$. 
This is due to the mass of the $q\bar q$ in the quark loop being $\propto Q^2$.
For such a $x_1/x_2$ the ratio of
nondiagonal and diagonal gluon densities weakly depends
on $x_2$. Hence with an accuracy of a few percent we can approximate
this ratio by its value  at $x_1/x_2=2$. Therefore, in the calculation of $R$,
 we used Eq.\ \ref{gluondis} and \ref{qdis} for both the 
diagonal and nondiagonal case but then multiplied the real photon result 
of the amplitude by a function $f(Q^2)$ for each $x$ and $Q^2$ to take 
into account the different evolution of the nondiagonal distribution as 
compared to the diagonal one,
\beq
ImA(x,Q^2,Q_0^2) = ImA(x,Q_0^2)+
\int^{Q^2}_{Q_0^2}{dQ'^{2}\over Q'^{2}}f(Q'^2) 
\int^{1}_{x} \frac{dx_1}{x_1}P_{qg}(x/x_1,\Delta/x_1)g(x_1,x_2,Q'^2).
\eeq
The function was determined by 
using our modified version of the CTEQ-package and, starting from the same 
initial distribution and evolving the diagonal and nondiagonal distribution
to a certain $Q^2$. We then0 compared the two distributions 
at the value $x_2=x_1/2=x$ for different $x$ and then interpolated for the
different ratios of the distribution in $Q^2$ for given $x$. 
For this median point the difference between the diagonal and nondiagonal 
gluon distribution is between $8\,- \, 25\%$ depending on the $x$ and $Q^2$ 
involved and $0\,-\,5\%$ for the quarks (see the figures in 
Ref.\ \cite{ours,av} for more details). 
  
As far as the complete amplitude at small $x$ is concerned, we can reconstruct
the real part via dispersion relations \cite{reim1,reim2}, which to a very 
good approximation gives:
\begin{equation}
\eta \equiv {Re A \over Im A}= \frac{\pi}{2}\frac{d\,\ln(x Im A)}{d\, 
\ln{1\over  x}}.
\label{reim}
\end{equation}
Meaning that  since $Im A$ can be fitted as $x^{-1-\delta}$, 
$\eta \approx {\pi \over 2}\delta$ 
is independent of $x$ to a good precision.
Therefore, our claims for the imaginary part of the 
amplitude also hold for the whole amplitude at small $x$.
This is due to the fact that within the dispersion representation of the 
amplitude over $x$ the contribution of the subtraction constant becomes 
negligible at sufficiently small $x$.

One also has to note that there is a potential pitfall since the
QED bremsstrahlung - the  Bethe-Heitler process, where the electron
interacts with a proton via a soft Coulomb photon exchange and the 
real photon is radiated off the electron, can be a considerable background. As
was shown by Ji \cite{6}, the Bethe-Heitler process will give a strong 
background at small $t$ and medium $Q^2$ and $x \ge 0.1$. We will discuss this
subject in more detail later on.

\section{The {\it \lowercase {t}}-slope of the $\gamma^*N\to \gamma N$ cross 
section}
\label{Bslope}
The slope of the differential cross section of the virtual Compton scattering
${d \sigma^{\gamma^*N\to \gamma N}\over dt} \propto \exp(Bt)$ is determined by 
three effects: (i) the average transverse size of the $q\bar q$ component of 
the $\gamma^*$ and $\gamma$ wave functions involved in the transition,
(ii) the pomeron-nucleon form factor in the nucleon vertex, and (iii)
Gribov diffusion in the soft part of the ladder.
This leads to several qualitative phenomena.
In the normalization point, $q\bar q $
configurations of an average transverse
size, comparable to that of the $\rho$-meson,
give the dominant contribution to the scattering amplitude,
 leading to a slope similar to that of the  processes 
$\gamma + p \to \rho, \omega + p$. 
The contribution of the higher mass $q \bar q$ components 
is known to result in an enhancement of the differential cross
section of the Compton scattering at $t=0$ by a factor $\approx 2$
as compared to the prediction of the vector meson dominance model with 
$\rho, \omega, \phi, J/\psi$ intermediate states, see
e.g. \cite{FNAL}.
 Since the diffraction of a photon to masses $M_X \ge 1.3 GeV$  has 
a smaller $t$ slope than for transitions
to $\rho$ and $\omega$, one could expect that
the high mass  contribution would lead to  
a t-slope of the Compton cross section  being
somewhat smaller than for the production of
$\rho, \omega$-mesons.  However direct experimental
comparison \cite{FNAL} of the slopes of
the  Compton scattering and the $\omega$-meson
photoproduction at $\left<E_{inc}^{\gamma}\right> \approx 100 GeV$
finds these slopes to be the same within the  experimental errors.
Using these data, we can estimate the slope
of the amplitude for diffractive photon production in DIS
at HERA energies but at moderate Q-i.e. in the normalization point as
\beq 
B(s,Q_0^2) = B_{Comp.Scatt.}(s_0)+ 
 2 \alpha'\ln({s\over s_0}),
\label{Brho}
\eeq
where $\alpha' =0.25 GeV^{-2}$, $s_0=200 GeV^2$, and 
$B_{Comp.Scatt.}(s_0)=6.9 \pm 0.3 GeV^{-2}$\cite{f9}.   
Hence for HERA energies $B(W=200 GeV,Q_0^2) \sim 10 GeV^{-2}$.

In another limit of large $Q^2$ and large enough $x$, say 
$x \sim 10^{-2}$, the dominant $q\bar q $ configurations
have small a transverse size and the upper vertex
does not contribute to the slope. Furthermore,
the perturbative contribution occupies most of the rapidity
interval and leaves no phase space for the soft Gribov diffusion. In
this case, the slope is given by the square of the two-gluon form 
factor of the nucleon which corresponds to $B=B_{ggN} \approx 
4 \div 5 GeV^{-2}$ \cite{1,2}.

An interesting situation emerges in the limit of large but fixed $Q^2$
when the energy starts to increase. In this case, the perturbative part of the 
ladder has the length  
$\sim \ln ({Q^2\over m_\rho^2 \kappa})$.
Here $\kappa=x/x_0$, where
$x_0$ is the fraction $x$ of the parent parton at a soft scale. For HERA 
kinematics $\kappa \sim 0.1-0.3$ for $Q^2 \sim 10-20 GeV^2$ and increasing 
with increasing $Q^2$.
This is consistent with the observation of an approximate factorization
for diffraction in the case of high masses ($M^2 \ge 100 GeV^2$, 
$M^2 \gg Q^2$) in the scattering of real and virtual photons observed at 
HERA \cite{H1}, namely 
\beq
{1 \over \sigma_{tot}(\gamma N)}{d \sigma(\gamma N\to XN)(W,M_X) 
\over dt dM^2}
\approx 
{1 \over \sigma_{tot}(\gamma^* N)}{d \sigma(\gamma^* N\to XN)(W,M_X) 
\over dt dM^2}.
\label{diffus}
\eeq
The observed slope for these processes is $B \sim 7 GeV^{-2}$ which is
consistent with the presence of a cone shrinkage at the rate
$\sim 2 \alpha'\ln(W^2/M^2)$
  as compared to the data at lower energies where
 smaller values of $W^2/M^2$ were probed.
Similarly we can expect that for virtual Compton scattering at large $Q^2$,
the slope will increase with decrease of $x$, at very small $x$,
 approximately as
\beq
B(W^2,Q^2)_{Q^2\gg \mu^2}=B_{ggN} + 2 \alpha'
(\ln(\frac{W^2 \kappa}{Q^2})-\ln (\frac{W_0^2}{m_{\rho}^2})
\theta(\frac{W^2 \kappa}{Q^2} - \frac{W_0^2}{m_{\rho}^2}),
\eeq
where $W_0^2=200 GeV^2$.  We take into account here that $B_{ggN}$
was determined experimentally from the processes at $W^2 \sim W_o^2$.

\section{The rate of exclusive photon production at HERA}
\label{sec:slope}

To check the feasibility of measuring a DVCS signal against the DIS 
background, 
we will be interested in the fractional number of DIS events
to diffractive exclusive photoproduction events at HERA in DIS given by:
\beq
R_{\gamma} = \frac{\sigma(\gamma^* +p \rightarrow \gamma + p)}
{\sigma_{tot}(\gamma^*p)}\simeq
\frac{d\sigma (\gamma^* + p \rightarrow \gamma +p)}{dt}|_{t=0}\times 
\frac{1}{B}/\sigma_{tot}(\gamma^*p),
\label{events}
\eeq
with 
\beq
\frac{d\sigma}{dt} (\gamma^* + p \rightarrow \gamma + p)= 
\frac{\sigma_{tot}^2(\gamma^*p)}{16\pi R^2}(1+\eta^2)
e^{Bt},
\label{N}
\eeq
from applying the optical theorem and where $R$ is the ratio of the amplitudes
given by Eq.\ \ref{ratio}, and $\eta= Re\,A/Im\,A$ is given by 
Eq.\ \ref{reim}, $t=-\frac{m_N^2 x^2}
{1-x}-p_t^2\simeq -p_t^2$ with $t_{min}=-\frac{m_N^2 x^2}{1-x}\simeq 0$.
Note, that even though we used the total DVCS cross section in 
Eq.\ \ref{events} we only need the ratio of the imaginary parts of the 
amplitudes to calculate $R_{\gamma}$. A complete expression for DVCS will be 
given in the next section. Note that only $d\sigma/dt(t=0)$
is calculable in QCD. The $t$ dependence is taken from data fits to hard 
diffractive processes.

Using the fact that $F_2(x,Q^2)\simeq \frac{\sigma_{tot}(\gamma^*p)Q^2}
{4 \pi^2 \alpha}$ one can rewrite Eq.\ \ref{events}:
\beq
R_{\gamma} \simeq \frac{\pi \alpha}{4 R^2 Q^2 B}F_2(x,Q^2)(1 + \eta^2).
\label{fN}
\eeq
where  $\eta^2 \simeq 0.09-0.27$ for the given 
$Q^2$ range. We computed $R_{\gamma}$, the fractional number of events given 
by Eq.\ \ref{fN}, for $x$ between
$10^{-4}$ and $10^{-2}$ and for a $Q^2$ of $2, 3.5, 12$ and $45\,\mbox{GeV}^2$ 
where the numbers for $F_2$ were taken from
\cite{15}. Based on our analysis of the previous section we use
Eq.\ \ref{Brho} for $Q^2=2~\mbox{GeV}^{2}$, assuming that for 
$Q^2=3.5~\mbox{GeV}^{-2}$ the slope drops by about 1 $\div $ 2 units as 
compared to Eq.\ \ref{Brho}
to account for the decrease of the transverse size of the $q\bar q$-pair;
for larger $Q^2$ we use Eq.\ \ref{diffus}.

We find $R_{\gamma}\simeq 1.1\times 10^{-3},\, 9.9\times 10^{-4}$ 
 at $x=10^{-4},\, 10^{-3}$ and $Q^2=2\mbox{GeV}^2$; $R_{\gamma}
\simeq 1.07\times 10^{-3},\, 9.3\times 10^{-4}$ 
  at $x=10^{-4},\, 10^{-3}$ and  $Q^2=3.5\mbox{GeV}^2$; 
$R_{\gamma}\simeq 4.5\times 10^{-4},\, 3.78\times 10^{-4}\, 
2.5\times 10^{-4}$ 
at $x=10^{-4},\, 10^{-3},\, 10^{-2}$ and $Q^2=12\mbox{GeV}^2$; and 
finally $R_{\gamma} \simeq 1.49\times 10^{-4},\, 1.04\times 10^{-4}$
at $x=10^{-3},\, 10^{-2}$ and $Q^2=45\mbox{GeV}^2$. 
As is to be expected, the number of events rises at small $x$ since the 
differential cross section is proportional to the square of the 
gluon distribution and the total cross section is just proportional to the 
gluon distribution i.e. the ratio in Eq.\ \ref{events} is expected to be 
proportional to the gluon distribution and this assumption is born out by
our calculation and falls with increasing $Q^2$ since $F_2$ does not grow as 
fast with energy.

\section{The complete cross section of exclusive photon production}
\label{sec:cross}

In order to study whether the Bethe-Heitler or DVCS Process will be dominant 
in real 
photon production we need the expressions for the differential cross sections 
first.

We find that the differential cross section for DVCS can be simply expressed 
through the DIS differential cross section by multiplying the
DIS differential cross section by $R_{\gamma}$ (see Eq.\ \ref{fN})
which was calculated in the previous section.
One can see this by observing how $F_2$ is related to 
$\sigma_{tot}(\gamma^*p)$ as given in Sec.\ \ref{sec:slope} and 
$\sigma_{tot}(\gamma^*p)$ to $\sigma_{DVCS}$ via $R_{\gamma}$ in the same 
section. We then find using Eq.\ \ref{fN} for $R_{\gamma}$
\beq
\frac{d\sigma_{DVCS}}{dxdyd|t|d\phi_r}=\frac{\pi\alpha^3s}{4R^2Q^6}(1+(1-y)^2)
e^{-B|t|}F^2_2(x,Q^2)(1+\eta^2)
\label{dvcs}
\eeq
with $\sigma_{DVCS} = \frac{d\sigma_{DVCS}}{dt}|_{t=0}
\times \frac{1}{B}$ using the same exponential $t$ dependence as in the 
previous section and $R$ being the ratio of the imaginary parts of the DIS to 
DVCS amplitudes as computed earlier.

\begin{figure}
\centering
\mbox{\epsfig{file=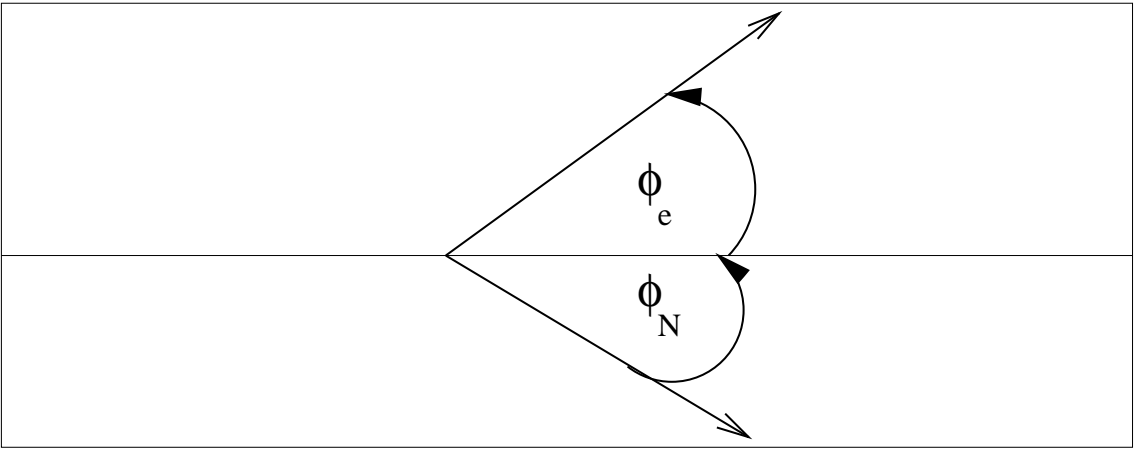,height=6cm}}
\vspace*{5mm}
\caption{The azimuthal final proton and electron angle in the transverse 
scattering plane.}
\label{angle}
\end{figure}

In writing Eq.\ \ref{dvcs} we neglected $F_L(x,Q^2)$ 
- the experimentally observed conservation of s channel helicities justifies
this approximation - and assumed $F_2\simeq 2xF_1$. $y=1-E'/E$ where $E'$ ist the energy of the 
electron in the final state and $\phi_r=\phi_N + \phi_e$, where $\phi_N$ is the 
azimuthal angle between the plane defined by 
$\gamma^*$ and the final state proton and the $x-z$ plane and $\phi_e$
is the azimuthal angle between the plane defined by the initial and final state
electron and the $x-z$ plane (see Fig.\ \ref{angle}).

In case of the Bethe-Heitler process, we find the differential cross section at
small $t$ to be
\bea
\frac{d\sigma_{BH}}{dxdyd|t|d\phi_r} &=& \frac{\alpha^3 s y^2(1+(1-y)^2)}{\pi Q^4
|t| (1-y)}\times \left [ \frac{G_E^2(t) + \tau G_M^2(t)}{1+\tau} \right ]
\label{difcros}
\eea
with $\tau = |t|/4m_N^2$, $s$ being the invariant energy and $y$ the fraction of 
the scattered electron/positron energy.
$G_E(t)$ and 
$G_M(t)$ are the electric and nucleon form factors and we describe them using 
the dipole fit
\beq
G_E(t)\simeq G_D(t)=(1+\frac{|t|}{0.71})^{-2}~~\mbox{and}~~G_M(t)=\mu_p G_D(t),
\label{nform}
\eeq
where  $\mu_p=2.7$ is the proton magnetic moment.  
We make the standard assumption that the spin flip term is small in the 
strong amplitude for small $t$.

In order to write down the complete total cross section of exclusive photon 
production we need the interference term between DVCS and Bethe-Heitler. Note
that in the case of the interference term one does not have a spinflip in the 
Bethe-Heitler amplitude, i.e.\ , one only has $F_1(t)$, as compared to 
Eq.\ \ref{difcros} containing a spinflip part, i.e.\ , $F_2(t)$. The 
appropriate combination of $G_E(t)$ and $G_M(t)$ which yields $F_1(t)$ is  
\beq
\left[\frac{G_E(t) + \frac{|t|}{4m_N^2}G_M(t)}{1+\frac{|t|}{4m_N^2}}\right ].
\eeq

We then find for the interference term of the differential cross section, 
where we already use Eq.\ \ref{difcros}, 
\bea
\frac{d\sigma_{DVCS+BH}^{int}}{dxdyd|t|d\phi_r} &=& \pm \frac{\eta 
\alpha^3 s y(1+(1-y)^2) cos(\phi_r) e^{-B|t|/2} F_2(x,Q^2)}{2 Q^5 \sqrt(|t|)
\sqrt(1-y) R}\nonumber\\
& & \times \left [ \frac{G_E(t) + \tau G_M(t)}{1+\tau} \right ]
\label{inter}
\eea 
with the + sign corresponding to electron scattering off a proton and the - 
sign corresponding to the positron. The total cross section is then just the 
sum of Eq.\ \ref{dvcs},\ref{difcros} and\ \ref{inter}.

\subsection{{\it \lowercase {t}}-dependence of Bethe-Heitler as compared
to DVCS for different $Q^2$}

\begin{figure}
\vskip-1in
\centering
\epsfig{file=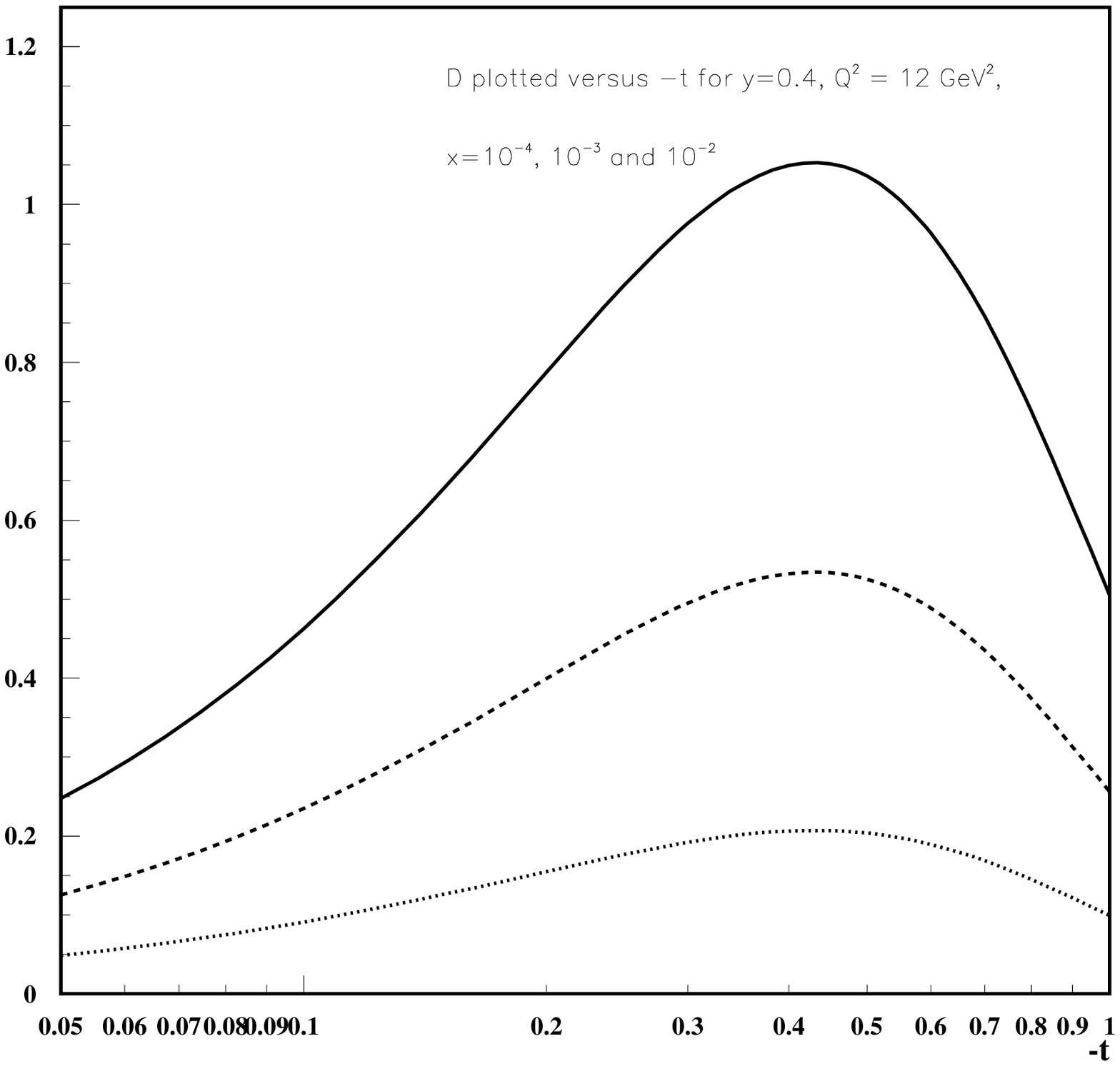,height=12cm}
\vskip-0.9in
\epsfig{file=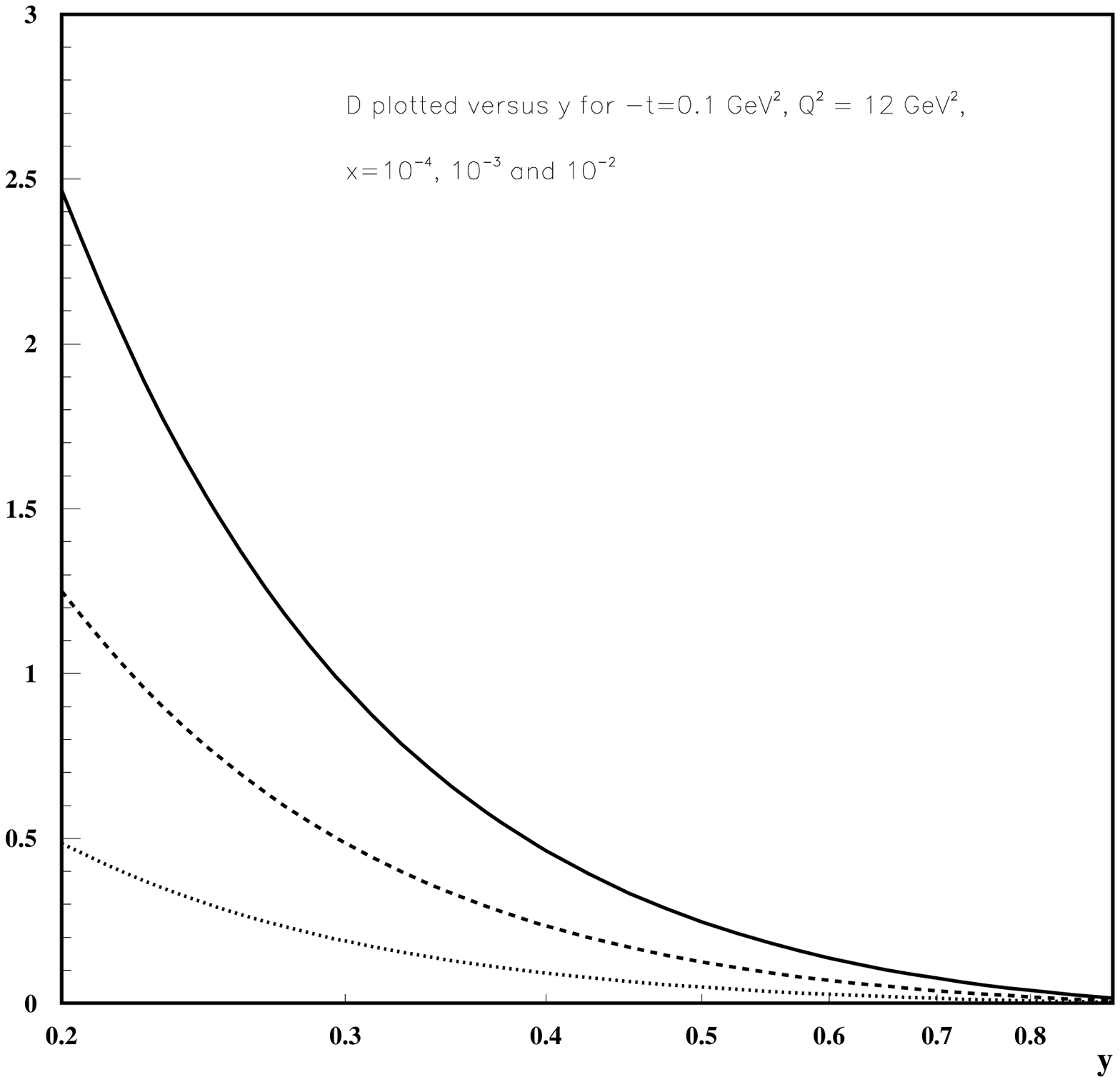,height=12cm}
\vskip-0.5in
\vspace*{5mm}
\caption{a) $D$ is plotted versus $-t$ for $x=10^{-4},10^{-3},10^{-2}$, 
$Q^2=12~\mbox{GeV}^2$, $B=5~\mbox{GeV}^{-2}$ and $y=0.4$. 
The solid curve is for $x=10^{-4}$, the dotted one for $x=10^{-2}$ and the 
dashed one for $x=10^{-3}$. b) $D$ is plotted versus $y$ for the same $x,Q^2,B$
and $-t=0.1~\mbox{GeV}^{2}$}
\label{tdep1}
\end{figure}

\begin{figure}
\vskip-1in
\centering
\epsfig{file=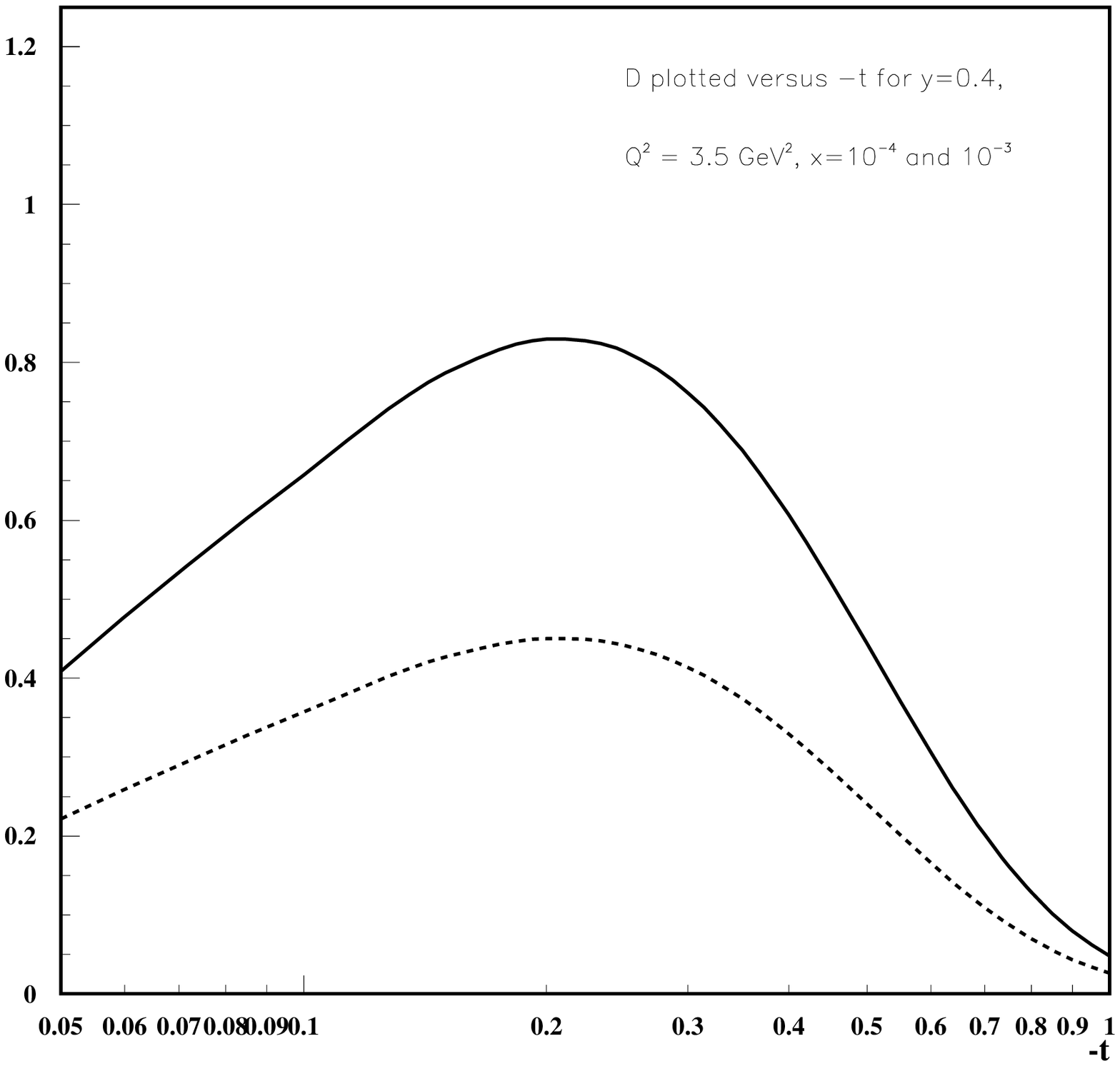,height=12cm}
\vskip-0.9in
\epsfig{file=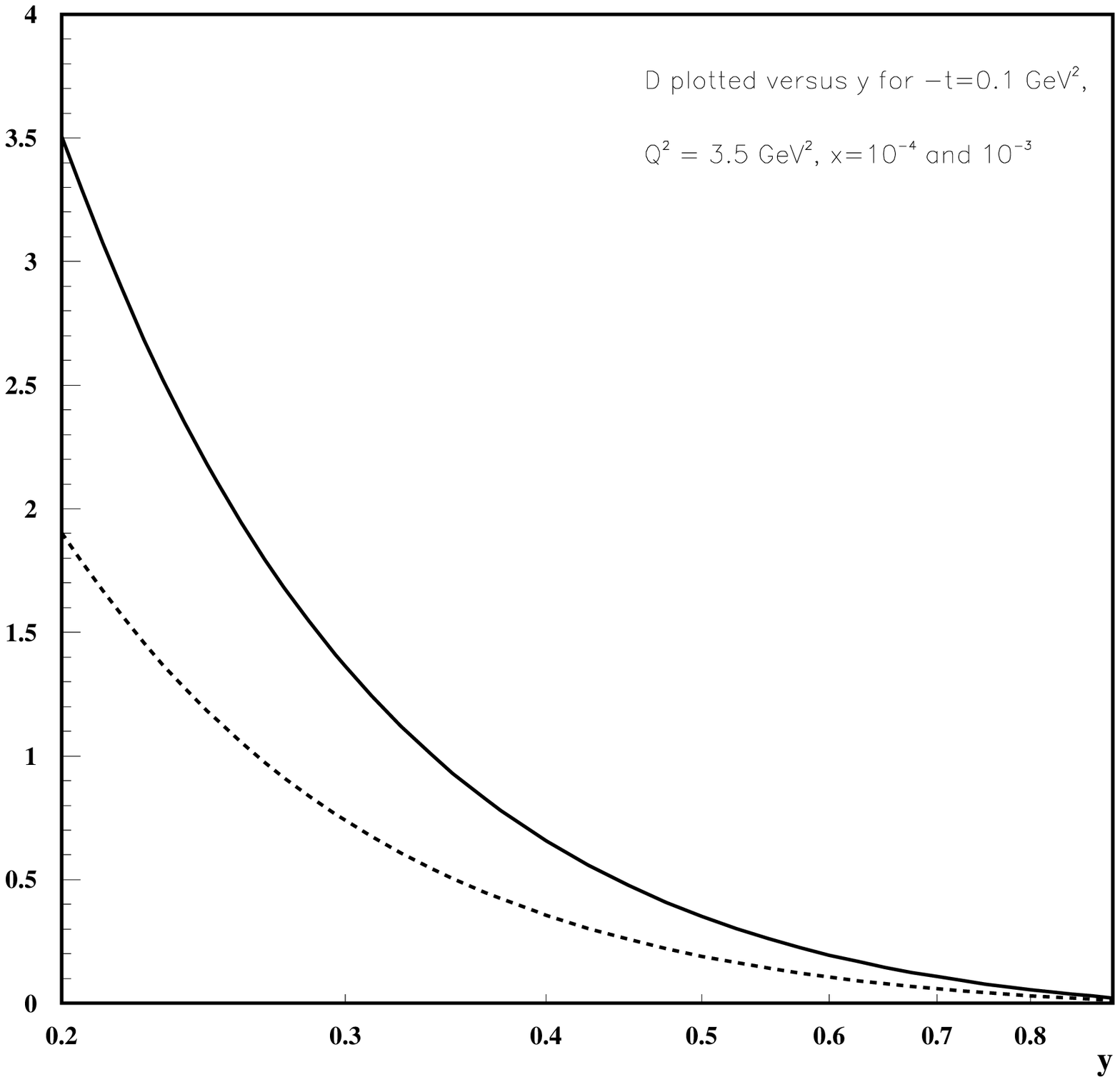,height=12cm}
\vskip-0.5in
\vspace*{5mm}
\caption{a) $D$ is plotted versus $-t$ for $x=10^{-4}$ and $10^{-3}$, 
$Q^2=3.5~\mbox{GeV}^2$, $B=8~\mbox{GeV}^{-2}$ and $y=0.4$. 
The solid curve is for $x=10^{-4}$, the dashed one for $x=10^{-3}$. 
b) $D$ is plotted versus $y$ for the same $x,Q^2,B$
and $-t=0.1~\mbox{GeV}^{2}$}
\label{tdep2}
\end{figure}

At this point it is important to determine how large the Bethe-Heitler 
background is as compared to DVCS for HERA kinematics,
hence, in the following 
discussion, we will estimate the ratio $D$ allowing a background comparison:
\beq
D=\frac{<d\sigma_{DVCS+BH}/dxdydt>}{<d\sigma_{BH}/dxdydt>}-1.
\eeq
with $<...>=\int_{0}^{2\pi}d\phi_r$. Using the expressions from 
Sec.\ \ref{sec:cross}, we compute $D$ and find $D > 1$ 
[see Figs.\ \ref{tdep1}a and \ref{tdep2}a] for relatively small $y$ and 
$0.1\leq t \leq 0.6$
with the given values of $x$ and $Q^2$ considered.
Note, however, that this does not mean that the case for DVCS is hopeless. 
As it turns out, it is rather advantageous to have $D<1$ when looking at the 
interference term which we will do next. Also note that there is a very 
strong energy dependence of $D$ which extends the range where DVCS is 
significantly larger than Bethe-Heitler with increasing energy.

\begin{figure}
\vskip-1in
\centering
\epsfig{file=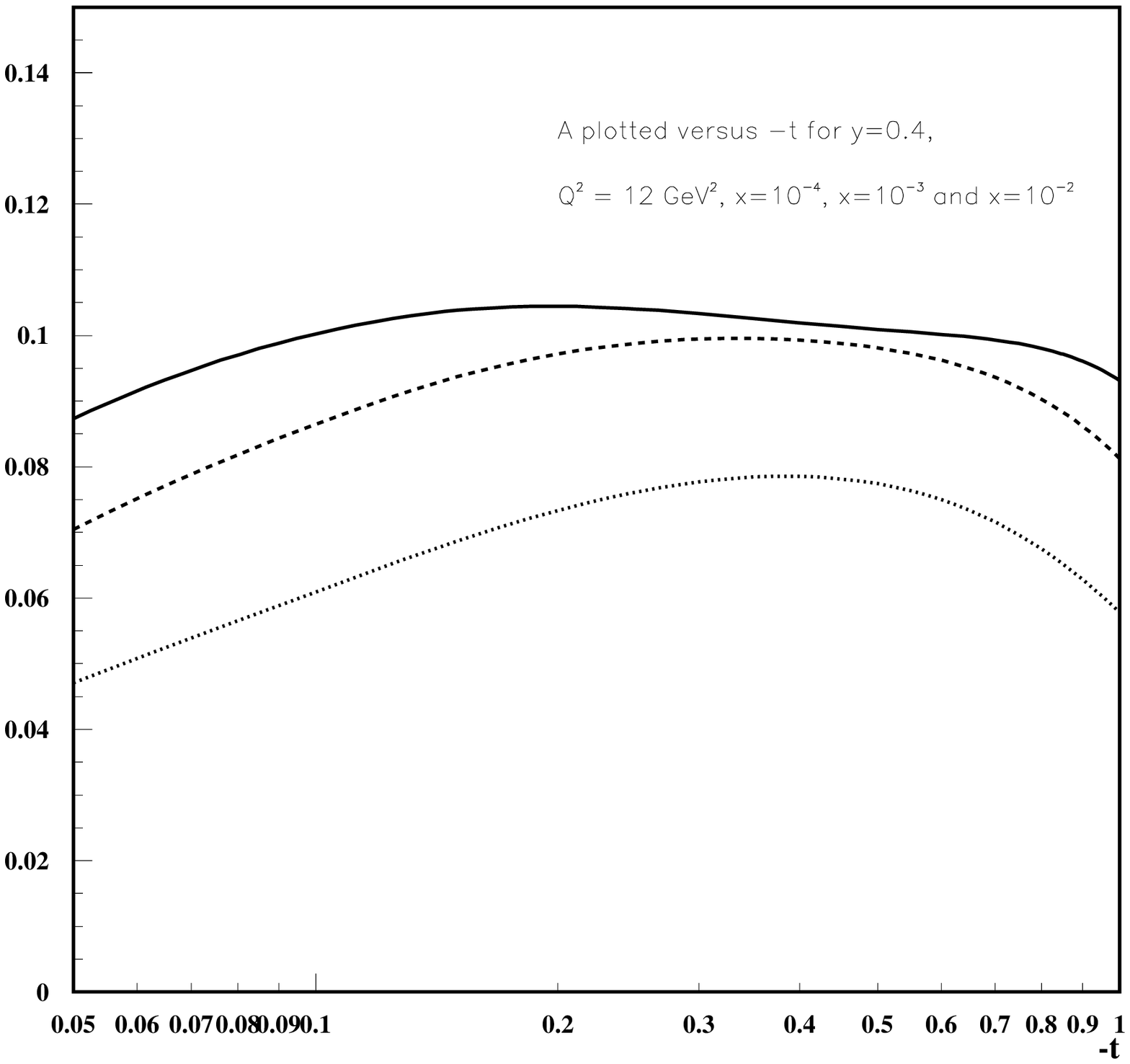,height=12cm}
\vskip-0.9in
\epsfig{file=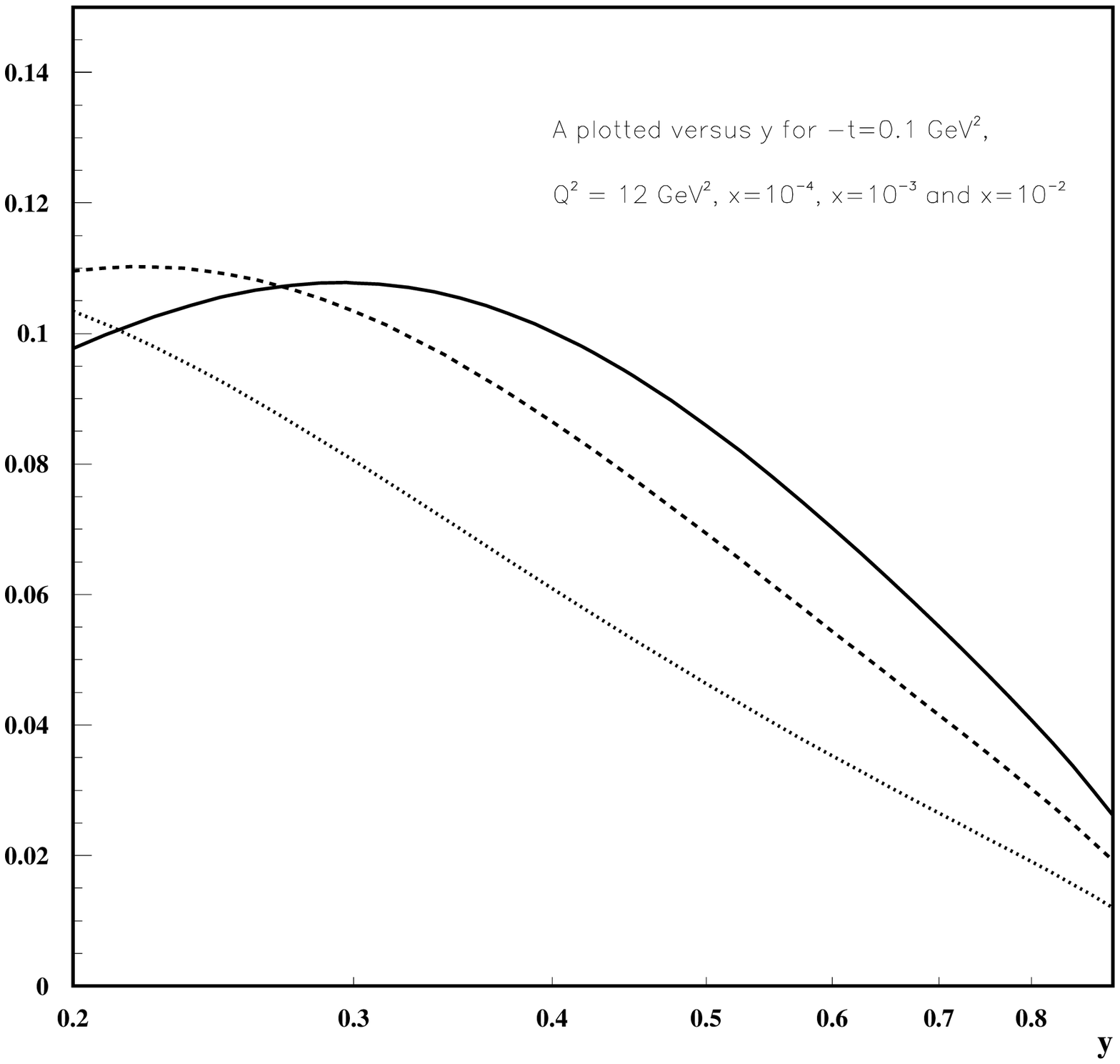,height=12cm}
\vspace*{5mm}
\vskip-0.5in
\caption{a) The asymmetry $A$ is plotted versus $-t$ for 
$x=10^{-4}$ (solid curve), 
$x=10^{-2}$ (dotted curve) and $x=10^{-3}$ (dashed curve)
again for $Q^2=12~\mbox{GeV}^2$, $B=5~\mbox{GeV}^{-2}$ and $y=0.4$.
b) $A$ is plotted versus $y$ for the same $x,Q^2,B$ and 
$-t=0.1~\mbox{GeV}^{2}$.}
\label{tdep4}
\end{figure}

\begin{figure}
\vskip-1in
\centering
\epsfig{file=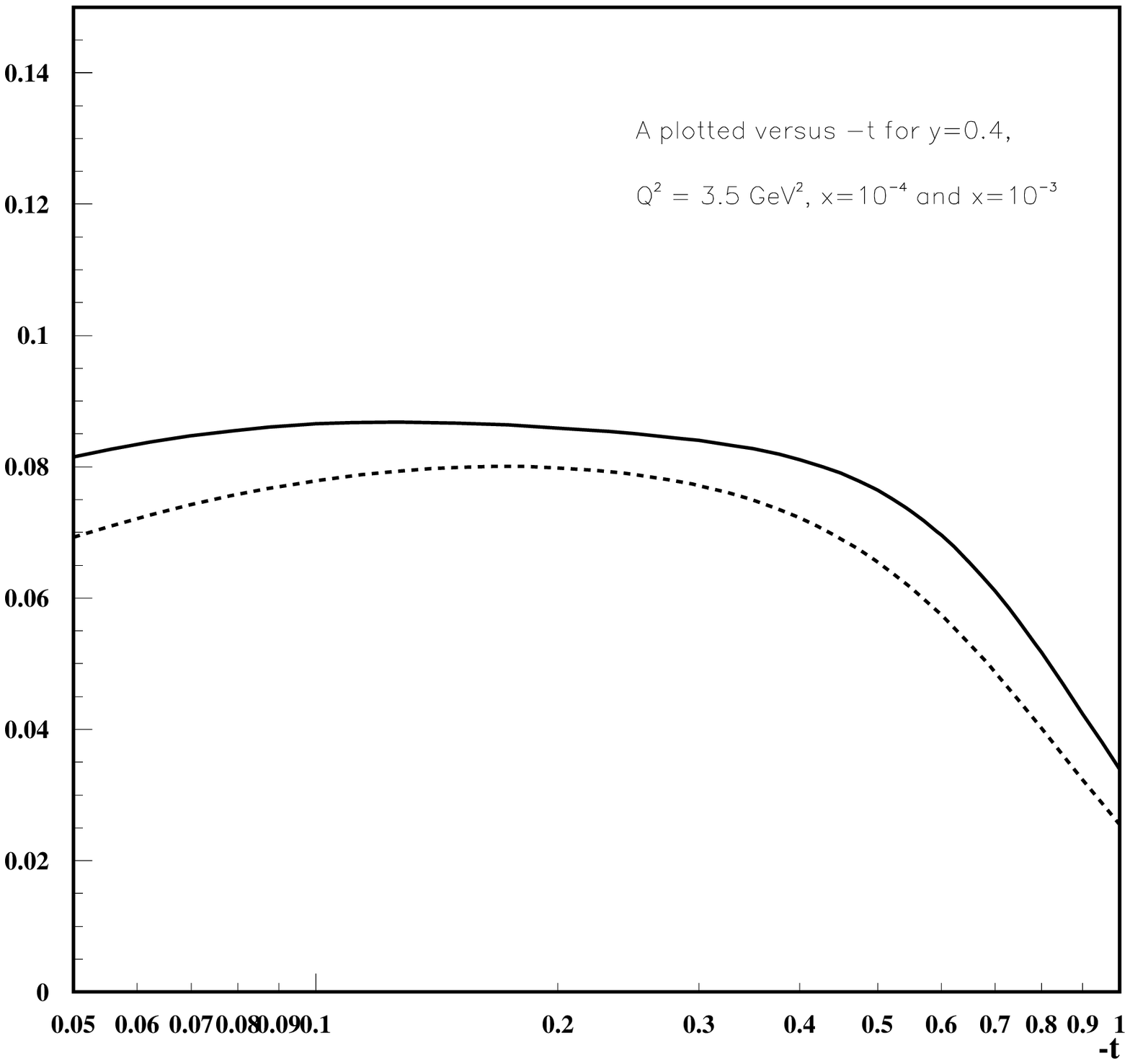,height=12cm}
\vskip-0.9in
\epsfig{file=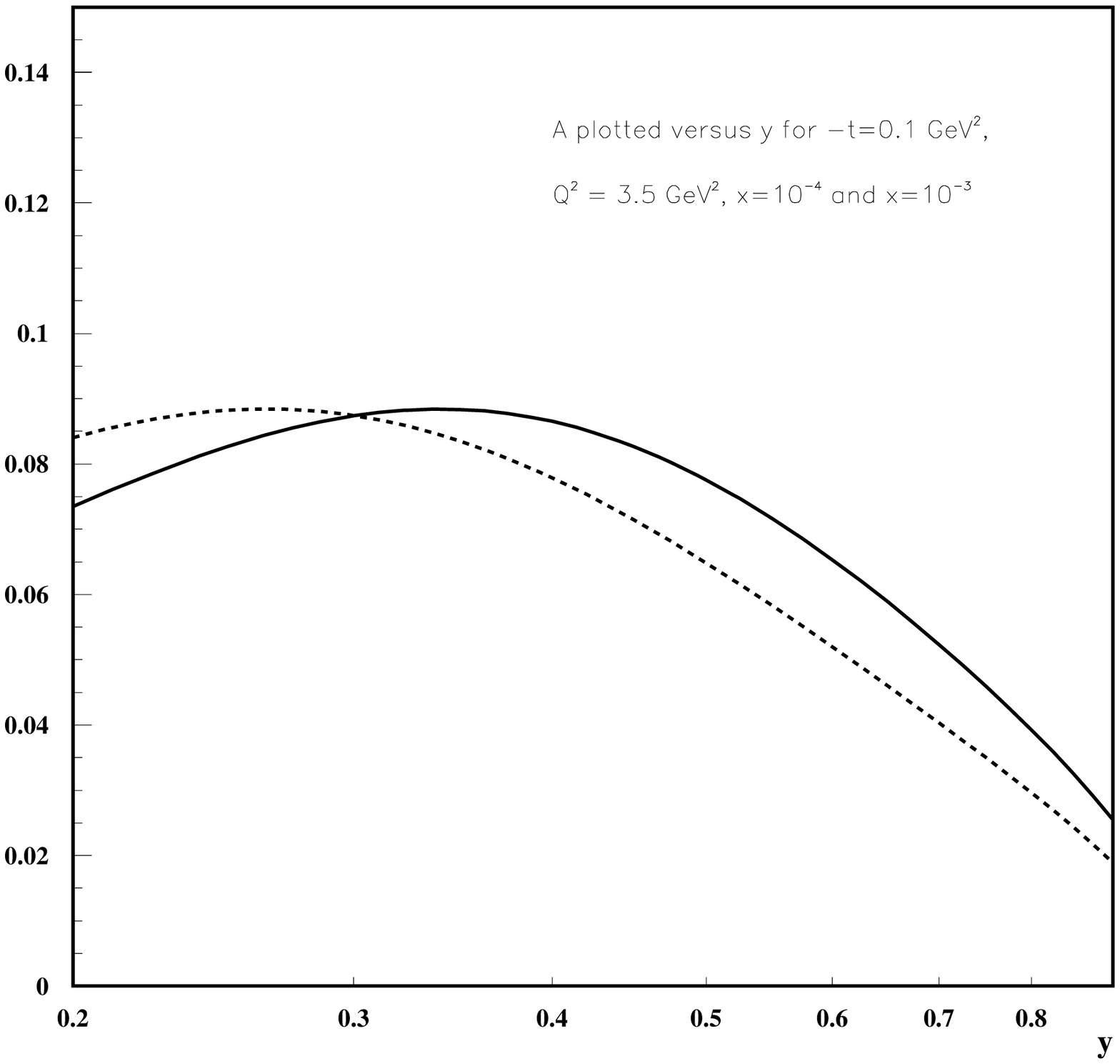,height=12cm}
\vskip-0.5in
\vspace*{5mm}
\caption{a) The asymmetry $A$ is plotted versus $-t$ for 
$x=10^{-4}$ (solid curve), 
$x=10^{-3}$ (dashed curve) again for $Q^2=3.5~\mbox{GeV}^2$, 
$B=8~\mbox{GeV}^{-2}$ and $y=0.4$.
b) $A$ is plotted versus $y$ for the same $x,Q^2,B$ and 
$-t=0.1~\mbox{GeV}^{2}$.}
\label{tdep3}
\end{figure}

It is convenient to illustrate the magnitude of the intereference term in 
the total cross section by considering the asymmetry for proton and either 
electron or positron to be in the same and opposite hemispheres ( we omit the 
rather cumbersome explicit expression but the reader can easily deduce it 
from Eq.\ \ref{dvcs},\ref{difcros} and\ \ref{inter}. )
\beq
A =\frac{\int_{-\pi/2}^{\pi/2}d\phi_r d\sigma_{DVCS+BH} - \int^{3\pi/2}_{\pi/2}
d\phi_r d\sigma_{DVCS+BH}}{\int_{0}^{2\pi}d\phi_r d\sigma_{DVCS+BH}}
\label{assym1}
\eeq
in other words, one is counting the number of events in the upper hemisphere 
of the detector minus the number of events in the lower half, normalized to 
the total cross section.
Fig.\ \ref{tdep4}a,b and \ref{tdep3}a,b show $A$ for the same kinematics as 
above and we find that the 
asymmetry is fairly sizeable already for small $t$ and is strongly dependent
on the energy. Due to this fairly large asymmetry, one has a first chance to 
access nondiagonal parton distributions through this asymmetry. 
We will discuss $A$ in more detail, in particular its energy dependence, in 
an upcoming paper.
 
Note, there is an increased experimental difficulty to measure DVCS if the 
recoil proton is not detected in other words if $t$ is not directly measured.
However there is a simple, practical way around this problem which we will 
discuss next.

\subsection{DVCS alternative to tagged proton in the final state}

Another interesting process, which can be studied in the context of DVCS,
 is the one where the nucleon dissociates into mass ``X''
 - $\gamma^* +p \to \gamma +X$. Perturbative QCD is applicable in this case 
as well. In particular the following factorization relation 
should be valid at sufficiently large $Q^2$:
\beq
\frac{\frac{d\sigma}{dt}(\gamma^* + p \rightarrow \gamma + X)}{\frac{d\sigma}
{dt}(\gamma^* + p \rightarrow \gamma + p)}\simeq
\frac{\frac{d\sigma}{dt}(\gamma^* + p \rightarrow J/\psi + X)}{\frac{d\sigma}
{dt}(\gamma^* + p \rightarrow J/\psi + p)}.
\label{ratio1}
\eeq
The big advantage of the dissociation process as compared to the process where
the target proton stays intact is that the Bethe-Heitler process is strongly 
suppressed for inelastic diffraction at small t due to the conservation of the 
electro-magnetic current, hence the amplitude is multiplied by an additional 
factor $\sqrt{\left|t\right|}$ which is basically $0$ for the Bethe-Heitler 
process. Thus, 
 the masking of the strong amplitude of photoproduction is small in this case.
Since there is already data available on $J/\psi$ production, this quantity 
can give us information on how different the slopes for the production of 
massless to massive vector particles are, providing us with more understanding
on how different or similar the exact production mechanisms are. Note that 
the ratio 
of the total dissociative to elastic cross section of $\rho$ meson production 
is found to be about $0.65$ at large $Q^2$ \cite{14} which is basically of 
$O(1)$. The same 
should hold true for $J/\psi$ production and in fact this ratio should be a 
universal quantity. This is due to the fact that one has complete 
factorization, hence the hard part plus vector meson is essentially a point 
and thus for the soft part, is does not matter what kind of vector
particle is produced.
The above said implies for Eq.\ \ref{ratio1} that it also should be of order 
unity, implying that the order of magnitude of the fractional number of events
for real photon production to DIS remains unchanged even though the actual 
number of $R_{\gamma}$ might decrease by as much as $35\%$. 
          
\section{Conclusions}
\label{sec:concl}

In the above said we have shown that pQCD is applicable to exclusive 
photoproduction by showing that the ratio of the imaginary parts of the 
amplitudes of DIS to a real photon is calculable in pQCD after specifying 
initial conditions since the derivative in energy of the hard scattering 
amplitudes can be unambiguously calculated in pQCD and all 
the non-perturbative physics can then be absorbed into a parton 
distribution. We wrote down an evolution equation for the imaginary part of 
the amplitude, which can be generalized to the complete amplitude at small $x$,
and solved for the imaginary part of the amplitude.
We also found that the imaginary part of the  amplitude
of the production of a real photon is larger than the one in 
the case of DIS in a broad range of $Q^2$ for the reasons as discussed above.
We also found the same to be true for the full amplitude at
small $x$. We also make experimentally testable predictions for
the number of real photon events and suggest that the number of events are 
small but not too small such that after improving the statistics on existing
or soon to be taken data, it would be 
feasible to access the nondiagonal gluon distribution at small $x$ from this
clean process. Finally, we demonstrated that measuring the asymmetry $A$ at HERA, which is 
fairly sizable in the kinematics in question, would allow one to determine the 
real part of the DVCS amplitude, in other words gain a first experimental 
insight into nondiagonal parton distributions, despite $D < 1$. 
     
\section*{Acknowledgments} 
We thank A.Mueller who back in 1995 drew our attention to the importance
of  this process.
We would like to thank A.V. Radyushkin for critical comments 
on an earlier draft of this paper and R. Engel
who drew our attention to the fact that H1 data indicate an independence 
of $Q^2$ of the ratio of inclusive high mass hadron production to the total
cross section of real and virtual photons.
This work was supported by the Department 
of Energy under grant number DE-FG02-93ER40771.



\begin{references}

\bibitem {1}  S.J. Brodsky, L.L. Frankfurt, J.F. Gunion, A.H. Mueller, 
              and M. Strikman, Phys. Rev. {\bf D50} (1994) 3134; 
              see also \cite{2}

\bibitem {2}  L.L. Frankfurt, W. Koepf, and M. Strikman, Phys. Rev. 
              {\bf D54} (1996) 3194.    

\bibitem {3}  A. Radyushkin Phys. Lett. {\bf B385} (1996) 333.

\bibitem {4}  J.C. Collins, L. Frankfurt, and M. Strikman, 
              Phys. Rev. {\bf D56} (1997) 2982.

\bibitem {ours} L.L. Frankfurt, A. Freund, V. Guzey and M. Strikman,
                Phys. Lett. {\bf B 418}, 345 (1998).

\bibitem {6}  X.-D. Ji, Phys. Rev. {\bf D55} (1997) 7114. .

\bibitem {7}  A. Radyushkin, hep-ph/9704207, A. Radyushkin, Phys.Lett 
              {\bf B380} (1996) 417 and private communication.

\bibitem {8}  I.I Balitsky and V.M. Braun, Nucl. Phys. {\bf B311}, 541 
              (1989).

\bibitem {9}  J. Bluemlein, B. Geyer and D. Robaschik, hep-ph/9705264.

\bibitem{MR97} A.Martin and M.Ryskin, hep-ph/9711371.

\bibitem{av} A. Freund, V. Guzey, hep-ph/9801388.

\bibitem {10} J. Bartels and M.Loewe, Z.Phys. {\bf C12}, 263 (1982).

\bibitem{f1} Clean in the sense that the wave function of a spatially small 
             size configuration within a real photon is better known as 
             compared to the wave functions of vector mesons thereby removing 
             a big theoretical uncertainty in the determination of the gluon 
             distribution.

\bibitem{Collins} J.C. Collins, D. E. Soper, and  G. Sterman, in PERTURBATIVE 
                  QUANTUM CHROMODYNAMICS, ed .A.Mueller, pp 1-91, World 
                  Scientific Publ.,1989.

\bibitem{ca} J.C. Collins, A. Freund, hep-ph/9801262.

\bibitem{bj} J. D. Bjorken  in Proceedings of  the International Symposium on 
             Electron and Photon Interactions at High Energies,p. 281--297, 
             Cornell (1971); J. D. Bjorken and J. B. Kogut,  Phys. Rev. D8 
             (1973) 1341.

\bibitem{FS88} L. L. Frankfurt and M. Strikman, Phys. Rep. 160~(1988)~235;
               Nucl.Phys. B316(1989) 340.

\bibitem{Gribov} V. N. Gribov, Sov. Phys. JETP 30~(1969)~709.

\bibitem {FRS} L. Frankfurt, A. V. Radyushkin, M. Strikman, Phys. Rev. 
               {\bf D55} (1997) 98.

\bibitem{FGS} L.Frankfurt, V.Guzey, M.Strikman, hep-ph/9712339.

\bibitem{f2}  In the case of the imaginary part of the amplitude which we 
              discuss at this point, one has $x_1 > \Delta >0$ and we can 
              treat the soft part as a parton distribution function (the DGLAP
              regime), whereas if $0<x_1<\Delta$ one would have the situation 
              of a distributional amplitude as first discussed by Radyushkin 
              \cite{3} which is governed by the Brodsky-Lepage evolution 
              equations.


\bibitem{reim1} V.N.Gribov and A.A.Migdal, Yad.Fiz. 8 (1968) 1002
                Sov.J.Nucl.Phys. 8 (1969) 583.

\bibitem{reim2} J.B.Bronzan, Argonne symposium on the Pomeron, 
                ANL/HEP-7327 (1973) p.33; J.B.Bronzan, G.L.Kane, and 
                U.P.Sukhatme, Phys.Lett. {\bf B 49}, 272 (1974). 

\bibitem {12} M.A. Shifman,A.L. Vainshtein and V.I. Zakharov 
              Nucl.Phys.{\bf B136},157 (1978), M. Glueck, E. Reya, 
              Phys.Lett. {\bf B83}, 98 (1979).

\bibitem {Abramowicz} H. Abramowicz, L.L. Frankfurt and M. Strikman, 
                      DESY-95-047, SLAC Summer Inst. 1994:539-574.

\bibitem{f3} In Ref.\ \cite{7} a similar equation was derived for the complete
             amplitude for larger $x\simeq 0.1$, where the quark distribution 
             dominates and one only needs the $P_{qq}$ kernel. Of course, at 
             sufficiently small $x$ the contribution of this term is 
             numerically small.

\bibitem{f4} Note that this expression is defined differently from the gluon 
             $\rightarrow$ quark splitting kernel as given in e.g. 
             Ref.\ \cite{7} by a factor of $1/x_1$ due to the fact that the 
             additional $x_1$ already appears in the convolution integral for 
             the $\ln Q^2$ derivative.

\bibitem{f5} This  effect will be taken into account in the actual numerical 
             calculation - see discussion below.

\bibitem{f6} The tensor structure which is the same in both cases, namely:
\beq
-g_{\mu\nu} + \frac{p_{\mu}q_{\nu} + q_{\mu}p_{\nu}}{pq} + 2x\frac{p_{\mu}
p_{\nu}}{pq},\nonumber
\eeq
             cancels out in the ratio!

\bibitem{f7} We found them to be around $10\%$ in the ratio $R$.

\bibitem{f8} We used the u-quark parametrization for all light quarks for 
             simplicity, which is surely unprobelamtic at small-$x$.


\bibitem{FNAL}        A.M.Breakstone et al., Phys.Rev.Lett. {\bf 47}~(1981)~
                      1778;{\bf 47}~(1981)~1782.

\bibitem{f9} Note that the data \cite{FNAL} can be equally well described by 
             the fit $d \sigma/dt \propto \exp(Bt)$ with 
             $B=6.9 \pm 0.3 GeV^{-2}$ and by the 
             $d \sigma/dt \propto \exp(8.9t+2.2t^2)$ fit.

\bibitem{H1}          H1-Collaboration (S.Aid et al.), Phys.Lett. {\bf B358}~
                      (1995)~412.

\bibitem {15}         H1-Collaboration (S.Aid et al.), hep-ex/9603004.

\bibitem {14}         H1-Collaboration (C. Adloff et al.), hep-ex/9705014.

\end{references}
\end{document}